\documentclass[12pt]{article}
\usepackage{amsmath,amssymb,amsfonts,amsthm}

\title{Two Dimensional Isotropic Harmonic Oscillator on a
 Time-dependent Sphere}

\author{ Ali Mahdifar$ ^1 $,  Behrouz Mirza$^{2}$, Rasoul Roknizadeh$^{3,4}$
\\
\footnotesize{1- Department of Physics, Faculty of Science,
Shahrekord University, Shahrekord, 88186-34141, Iran} \\
\footnotesize{2- Department of Physics, Isfahan University of Technology, Isfahan, 84156-83111, Iran} \\
\footnotesize{3- Department of Physics, Faculty of Science, University of Isfahan, Hezar Jerib, Isfahan, 81746-73441, Iran}\\
\footnotesize{4- Physics Department, Stockholm University,
Stockholm, 10691, Sweden}
\\ \footnotesize{E-mails: mahdifar$ _{-} $a@sci.sku.ac.ir; b.mirza@cc.iut.ac.ir; rokni@sci.ui.ac.ir }}

\begin{document}

\maketitle

\begin{abstract}
In this paper, we investigate a two dimensional isotropic harmonic
oscillator on a time-dependent spherical background. The effect
of the background can be represented as a minimally coupled field
to the oscillator's Hamiltonian. For a fluctuating background,
transition probabilities per unit time are obtained. Transitions
are possible if the energy eigenvalues of the oscillator $E_i$
and frequencies of the fluctuating background $\omega_n$ satisfy
the following two simple relations: $E_{j}\simeq
E_{i}-\hbar\omega_{n}$ (stimulated emission) or $E_{j}\simeq
E_{i}+\hbar\omega_{n}$ (absorption). This indicates that a
background fluctuating at a frequency of $\omega_n$ interacts
with the oscillator as a quantum field of the same frequency. We
believe this result is also applicable for an arbitrary
quantum system defined on a fluctuating maximally symmetric
background.

 \end{abstract}

 {\bf Pacs: 03.65.-w, 04.62.+v}

 \maketitle
 %==========================================================================
 %==========================================================================

\newcommand{\be}{\begin{equation}}
\newcommand{\ee}{\end{equation}}
\newcommand{\bea}{\begin{eqnarray}}
\newcommand{\eea}{\end{eqnarray}}
\newcommand{\h}{\hspace{0.30 cm}}
\newcommand{\vs}{\vspace{0.30 cm}}
\newcommand{\n}{\nonumber}
%----------------------------------------------------------------------------------

\section{Introduction}

Higgs and Leemon investigated the non-relativistic motion of a
particle on a $N$-dimensional fixed sphere (embedded in the
Euclidean ($N+1$)-dimensional space) under conservative central
potentials \cite{Higgs},\cite{Leemon}. These central potentials
reduce to the familiar Coulomb and isotropic oscillator
potentials of an Euclidean geometry when the curvature of the
sphere goes to zero. Higgs defined the motion on a sphere of a
constant radius by means of a central (or gnomonic) projection
from the motion of a $N$-dimensional plane, tangent to the sphere
at a given point. The advantage of this projection over all
others for the analysis of particle motion  on a sphere derives
from the fact that free particle motion (uniform motion on a
great circle) projects into a rectilinear plane, while its
non-uniform motion projects into the tangent plane. In other
words, the projected free particle orbits are the same as those
in the Euclidean geometry: the curvature affects only the speed of
the projected motion. Higgs also showed that this feature persists
in the presence of a central force derived from a potential
$V(r)$; i.e., he proved that the dynamical symmetries in sphere
are the same as those in the plane. Using the Hamiltonians  as
functions of the Casimir operators, the eigenstates and
eigenvalues of the two systems were also obtained in \cite{Higgs}
and \cite{Leemon} respectively.

Recently, by exploring the Higgs model, we have constructed  the
generalized (Nonlinear) coherent states \cite{star3} of a 2D
harmonic oscillator constrained to a spherical surface, to study
some of their quantum optical properties \cite{xx}. Accordingly,
we have developed a feasible physical model to generate nonlinear
coherent states on a sphere in a generalized trapped ion system
\cite{AVogel}.

The geometry of the sphere coherent states  is studied  in
\cite{ijm}. It has been  shown that the structure and  properties
of sphere- (nonlinear) coherent states, could be explored to
studying the curvature effects of physical space on both
transition probability and  geometric phase. It has been shown,
the transition probability decreases with increasing  curvature
of the physical space which, in turn, affects  the geometric
phase.

In the present contribution, our main purpose is investigating  the
behavior of  an isotropic oscillator on a dynamical background.
For this purpose, we study a two dimensional isotropic harmonic
oscillator constrained to a spherical surface with a non-constant
(time-dependent) radius. Our ambition is  to incorporate the
variation of the sphere's radius in the isotropic oscillator's
Hamiltonian as a minimally coupled field. This model could be
related to many of the above mentioned topics and we will study
them in the future.

In next stage, by using the perturbation theory, we will study
the transition rate  between energy levels to achieve  a golden
rule.  Appreciable probabilities for transition are possible only
if the transition and radius fluctuation frequencies are nearly
on resonance, e.g.,  if ,$E_{f}\simeq E_{i}-\hbar\omega_{n}$
(stimulated emission) or by $E_{f}\simeq E_{i}+\hbar\omega_{n}$
(absorption). Thus,  the isotropic harmonic oscillator on the
sphere will be excited via  appropriate frequencies  of the
background fluctuations. The proposed  golden
rule is expected to be applicable for any quantum system defined in a
fluctuating maximally symmetric background. We expect that for a
general time dependent background, the golden rule is still valid
where   $\omega_{n}$'s are eigenmodes of the oscillation of the
time dependent background. This could also have interesting
applications in cosmology and opto-mechanical systems.

 \section{Isotropic Oscillator on a time-dependent Background}\label{geometric}

In this section, we obtain the Hamiltonian of a two dimensional
isotropic harmonic oscillator confined at a spherical  surface
of  non-constant (time-dependent) radius. For this purpose, we
use the gnomonic projection, which is the projection onto the
tangent plane from the centre of the sphere in the embedding
space. The advantage of this projection over all others for the
analysis of the motion of a particle on a sphere stems from the
fact that free particle motion (uniform motion on a great circle)
projects into rectilinear, but non-uniform motion on the tangent
plane. That is, the projected free particle orbits are the same
as in Euclidean geometry: the curvature affects only the speed of
the projected motion. This feature persists in the presence of a
central force derived from a potential $V(r)$ \cite{Higgs}. We
first obtain the metric of the time-dependent background in the
gnomonic coordinates and then calculate the Hamiltonian of the
harmonic oscillator.

Let us designate the Cartesian coordinates of the time-dependent
spherical background by $(q_{0},q_{1},q_{2})$, and assume that they satisfy the sphere equation all the time,
 \be
   q_{0}^{2}(t)+q_{1}^{2}(t)+q_{2}^{2}(t)=R^{2}(t),
 \ee
where, $R(t)$ is the radius of the sphere.

If we denote the Cartesian coordinates of the tangent plane to the
sphere by $x$ and $y$, the relationship between these two coordinates is given by
 \bea
   q_{1}&=&\frac{x}{\sqrt{[1+\lambda(x^{2}+y^{2})}]},\n\\
   q_{2}&=&\frac{y}{\sqrt{[1+\lambda(x^{2}+y^{2})}]},\n\\
   q_{0}&=&\pm\frac{1}{\sqrt{\lambda[1+\lambda(x^{2}+y^{2})}]},
 \eea
where, $\lambda=\frac{1}{R^{2}(t)}$ is the time-dependent
curvature of the sphere and $+$ ($-$) in the last line means the
upper (lower) half of the sphere. Accordingly, a point on the
sphere can be represented as,
 \be
   \vec{r}=(\frac{x}{\Lambda},\frac{y}{\Lambda},\frac{1}{\sqrt{\lambda}\Lambda}),
 \ee
where,
 \be
   \Lambda=\sqrt{1+\lambda (x^{2}+y^{2})}.
 \ee
Now the differential of $\vec{r}$ is given by,
 \be
   d \vec{r}=\vec{r}_{x} dx+\vec{r}_{y} dy+\vec{r}_{t} dt,
 \ee
where, $\vec{r}_{x}, \vec{r}_{y}$ and $\vec{r}_{t}$ are partial
derivatives of $\vec{r}$ with respect to $x, y$ and $t$,
respectively. Thus, after a straightforward calculation, we obtain the
metric of the sphere as,
 \bea\label{ds}
   ds^{2}
   &\doteq&
   d\vec{r}\cdot d\vec{r}=\frac{(\vec{x}\cdot
   d\vec{x})^{2}}{r^{2}\Lambda^{2}}+\frac{1}{\Lambda}[d\vec{x}\cdot d\vec{x}-\frac{1}{r^{2}}(\vec{x}\cdot
   d\vec{x})^{2}]\n\\
   &+&
   2(\vec{r}_{t}\cdot\vec{r}_{x})dt dx+2(\vec{r}_{t}\cdot\vec{r}_{y})dt
   dy+2(\vec{r}_{t}\cdot\vec{r}_{t})dt^{2},
 \eea
where,
 \be
   \vec{x}=(x,y),\quad r^{2}=x^{2}+y^{2}.
    \ee
Now, we can obtain $\dot{s}$ from equation
(\ref{ds}) as follows
 \bea\label{sdot}
   \dot{s}^{2}
   &=&
   \frac{(\vec{x}\cdot \dot{\vec{x}})^{2}}{r^{2}\Lambda^{2}}+\frac{1}{\Lambda}[\dot{\vec{x}}\cdot
   \dot{\vec{x}}-\frac{1}{r^{2}}(\vec{x}\cdot
   \dot{\vec{x}})^{2}]\n\\
   &+&
   2(\vec{r}_{t}\cdot\vec{r}_{x})\dot{x}+2(\vec{r}_{t}\cdot\vec{r}_{y})\dot{y}
   +2(\vec{r}_{t}\cdot\vec{r}_{t})
 \eea
In Ref. \cite{Higgs}, the authors showed that the Lagrangian of a
free particle on a sphere is affected by the curvature of the
sphere. By studying the dynamical symmetries of the sphere, they
also found that the form of the central potentials (such as
isotropic harmonic oscillator and Coulomb potentials) is not
affected by the curvature of the sphere if we use gnomonic
coordinates. We can, therefore, write the Lagrangian (using it's
standard definition) of a two dimensional isotropic oscillator on
a time-dependent spherical background by using the equation
(\ref{sdot}) as,
 \bea\label{Lag}
   {\cal{L}}
   &\doteq&
   \frac{1}{2}\dot{s}^{2}-V(x,y)=
   \frac{1}{2}\frac{(\vec{x}\cdot \dot{\vec{x}})^{2}}{r^{2}\Lambda^{2}}+\frac{1}{2 \Lambda}[\dot{\vec{x}}
   \cdot \dot{\vec{x}}-\frac{1}{r^{2}}(\vec{x}\cdot
   \dot{\vec{x}})^{2}]\n\\
   &+&
   (\vec{r}_{t}\cdot\vec{r}_{x})\dot{x}+(\vec{r}_{t}\cdot\vec{r}_{y})\dot{y}
   +(\vec{r}_{t}\cdot\vec{r}_{t})-V(x,y)
 \eea
where, $V(x,y)$ is the potential of the two-dimensional isotropic
oscillator on the tangent plane coordinates (we assume $ m = \omega
= 1)$,
 \be
  V(x,y)=\frac{1}{2}(x^{2}+y^{2}).
 \ee
By using the Lagrangian (\ref{Lag}), we can calculate the momentum
vector as,
 \bea
   \vec{p}
   &=&
   \frac{\partial {\cal{L}}}{\partial \dot{x}}\hat{i}+\frac{\partial {\cal{L}}}{\partial
   \dot{y}}\hat{j}\n\\
   &=&
   \frac{\vec{x}(\vec{x}\cdot \dot{\vec{x}})}{r^{2}\Lambda^{2}}
   +\frac{1}{\Lambda}[\dot{\vec{x}}-\frac{1}{r^{2}}(\vec{x}\cdot
   \dot{\vec{x}})]+\vec{A}(t)
 \eea
where,
 \be
   \vec{A}(t)=(\vec{r}_{t}\cdot\vec{r}_{x})\hat{i}+(\vec{r}_{t}\cdot\vec{r}_{y})\hat{j}.
 \ee
With proper calculation, we get  the Hamiltonian of the two-dimensional isotropic
oscillator constrained to  a time-dependent spherical background,
 \be\label{hclass}
   H\doteq
   \dot{\vec{x}}\cdot{\vec{p}}-{\cal{L}}=H_{0}(\vec{x},\vec{p}-\vec{A}(t);\lambda)+\phi(t),
 \ee
where, $\phi(t)=-(\vec{r}_{t}\cdot\vec{r}_{t})$ and
$H_{0}(\vec{x},\vec{p};\lambda)$ is the usual Hamiltonian of a
two-dimensional isotropic oscillator on a sphere obtained by
Higgs \cite{Higgs},
 \be
   H_{0}(\vec{x},\vec{p};\lambda)=\frac{1}{2}(\pi^{2}+\lambda L^{2})
   +\frac{1}{2}(x^{2}+y^{2}),
 \ee
where,
 \be\label{pi}
   \vec{\pi}=\vec{p}+\lambda\ \vec{x}(\vec{x}\cdot\vec{p}),
 \ee
and
 \be\label{l}
   \vec{L}=\vec{x}\times\vec{p}.
 \ee
It is obvious from equation (\ref{hclass}) that the effect of
a time-dependent spherical background appears in the related
Hamiltonian as a minimally coupled field.
Thus, we may interpret $\vec{A}(t)$ and $\phi(t)$ in a similar way as  the vector
and scalar potentials of an electromagnetic field if
we choose $e=c=1$.

To consider the quantum harmonic oscillator on a time-dependent
background, we quantize the Hamiltonaian (\ref{hclass}) by
replacing classical position and momentum by related operators,
 \be\label{hquantum}
   \hat{H}=\hat{H}_{0}(\vec{\hat{x}},\vec{\hat{p}}-\vec{\hat{A}}(t);\lambda)
   -(\vec{\hat{r}}_{t}\cdot\vec{\hat{r}}_{t}),
 \ee
where,
 \be\label{QHiggs}
             \hat{H}_{0}(\vec{\hat{x}},\vec{\hat{p}};\lambda)=\frac{1}{2}(\hat{\pi}^{2}+\lambda \hat{L}^{2})
             +\frac{1}{2}(\hat{x}^{2}+\hat{y}^{2}),
 \ee

  \noindent $\hat{\pi}$ and $\hat{L}$ are symmetric forms of (\ref{pi})
and (\ref{l}) given from the following two equations
    \begin{equation}\label{higgs 14a}
            \vec{\hat{\pi}}=\vec{\hat{p}}+\frac{\lambda}{2}\left[\vec{\hat{x}}(\vec{\hat{x}}.\vec{\hat{p}})
            +(\vec{\hat{p}}.\vec{\hat{x}})\vec{\hat{x}}\right],
    \end{equation}
and
            \begin{equation}\label{higgs 5.5}
            \hat{L}^{2}=\frac{1}{2}\hat{L}_{ij}\hat{L}_{ij},\h \hat{L}_{ij}
            =\hat{x}_{i}\hat{p}_{j}-\hat{x}_{j}\hat{p}_{i}.
            \end{equation}

 \section{Harmonic Oscillator on a Fluctuating Background}

In this section, we consider a two dimensional isotropic harmonic
oscillator on a fluctuating spherical background. The time
dependent radius of the sphere is defined by,
 \be
   R(t)=R_{0}+\sum_{n}\alpha_{n}\sin(\omega_{n}t),
 \ee
where, $R_{0}$ and $\alpha_{n}$'s are real constants and
$\omega_{n}$'s are frequencies of different modes of the
fluctuations. We also assume that the fluctuation amplitudes
$\alpha_{n}$ are very small compared to $R_{0}$ such that
$\{\forall n:\ \frac{\alpha_{n}}{R_{0}}\ll 1\}$. Therefore, we
can expand the curvature of the sphere as,
 \be
   \lambda\doteq\frac{1}{R^{2}}=\lambda_{0}
   \left[1-2(\lambda_{0})^{\frac{1}{2}}\sum_{n}\alpha_{n}\sin(\omega_{n}t)\right]
   +O(\lambda_{0}^{2}\alpha_{n}^{2}),
 \ee
where, $\lambda_{0}=\frac{1}{R_{0}^{2}}$ and
$O(\lambda_{0}^{2}\alpha_{n}^{2})$ means the terms of order equal
to or greater than  $\lambda_{0}^{2}\alpha_{n}^{2}$, which are
negligible for this system. According to this limiting case, we
also obtain
 \bea
   \phi(t)
   &\cong&
   0,\n\\
   \vec{\hat{A}}(t,\vec{\hat{x}})
   &\cong&
   f(t)\
   \vec{\hat{m}}(\vec{\hat{x}}),
 \eea
where,
 \be
   f(t)=-(\lambda_{0})^{\frac{1}{2}}\sum_{n}\alpha_{n} \omega_{n}\cos(\omega_{n}t),
 \ee
and
 \be
   \vec{\hat{m}}(\vec{\hat{x}})=\frac{\vec{\hat{x}}}{\left[1+\lambda_{0}
   (\vec{\hat{x}}.\vec{\hat{x}})\right]^{2}}.
 \ee
Thus, from equation (\ref{hquantum}), the Hamiltonian of the
two-dimensional harmonic oscillator on a fluctuating spherical
background can be written as,
 \be\label{ht}
   \hat{H}\cong\hat{H}_{0}(\vec{\hat{x}},\vec{\hat{p}};\lambda_{0})
   +\hat{V}(t,\vec{\hat{x}},\vec{\hat{p}};\alpha;\lambda_{0}),
 \ee
where, the time-independent Higgs Hamiltonian,
$\hat{H}_{0}(\vec{\hat{x}},\vec{\hat{p}};\lambda_{0})$, is given
by equation (\ref{QHiggs}) and the time-dependent part
$\hat{V}(t,\vec{\hat{x}},\vec{\hat{p}};\alpha;\lambda_0)$ is
defined as,
 \be\label{vt}
   \hat{V}(t,\vec{\hat{x}},\vec{\hat{p}};\alpha;\lambda_{0})=V_{0}(t;\alpha)
   \hat{V}_{1}(\vec{\hat{x}},\vec{\hat{p}};\lambda_{0})+\widetilde{V}_{0}(t;\alpha)
   \hat{\widetilde{V}}_{1}(\vec{\hat{x}},\vec{\hat{p}};\lambda_{0}),
 \ee
where,
 \be\label{v0}
   V_{0}(t;\alpha)=(\lambda_{0})^{\frac{1}{2}}\sum_{n}\alpha_{n}
   \omega_{n}\cos(\omega_{n}t),
 \ee
 \be\label{vp0}
   \widetilde{V}_{0}(t;\alpha)=(\lambda_{0})^{\frac{1}{2}}\sum_{n}\alpha_{n}
   \sin(\omega_{n}t),
 \ee
 \bea
   \hat{V}_{1}(\vec{\hat{x}},\vec{\hat{p}};\lambda_{0})
   &=&
   \vec{\hat{p}}.\vec{\hat{m}}+\vec{\hat{m}}.\vec{\hat{p}}\n\\
   &+&
   \frac{\lambda_{0}}{2}\left\{\vec{\hat{p}}.\left[\vec{\hat{x}}(\vec{\hat{x}}.\vec{\hat{m}})
   +(\vec{\hat{m}}.\vec{\hat{x}})\vec{\hat{x}}\right]
   +\vec{\hat{m}}.\left[\vec{\hat{x}}(\vec{\hat{x}}.\vec{\hat{p}})
   +(\vec{\hat{p}}.\vec{\hat{x}})\vec{\hat{x}}\right]\right\}\n\\
   &+&
   \lambda_{0}\left[(\hat{x}\hat{p}_{y}-\hat{y}\hat{p}_{x})(\hat{x}\hat{m}_{y}-\hat{y}\hat{m}_{x})
   +(\hat{x}\hat{m}_{y}-\hat{y}\hat{m}_{x})(\hat{x}\hat{p}_{y}-\hat{y}\hat{p}_{x})\right]\n\\
   &+&
   \frac{\lambda_{0}^{2}}{2} \{[\vec{\hat{x}}(\vec{\hat{x}}.\vec{\hat{p}})+(\vec{\hat{p}}.\vec{\hat{x}})\vec{\hat{x}}].
   [\vec{\hat{x}}(\vec{\hat{x}}.\vec{\hat{m}})+(\vec{\hat{m}}.\vec{\hat{x}})\vec{\hat{x}}]\n\\
   &+&
   [\vec{\hat{x}}(\vec{\hat{x}}.\vec{\hat{m}})+(\vec{\hat{m}}.\vec{\hat{x}})\vec{\hat{x}}].
   [\vec{\hat{x}}(\vec{\hat{x}}.\vec{\hat{p}})+(\vec{\hat{p}}.\vec{\hat{x}})\vec{\hat{x}}]\},
 \eea
and,
 \bea
   \hat{\widetilde{V}}_{1}(\vec{\hat{x}},\vec{\hat{p}};\lambda_{0})
   &=&
   -\left\{L^{2}+\frac{1}{2}\vec{\hat{p}}.\left[\vec{\hat{x}}(\vec{\hat{x}}.\vec{\hat{p}})
   +(\vec{\hat{p}}.\vec{\hat{x}})\vec{\hat{x}}\right]+\frac{1}{2}\left[\vec{\hat{x}}(\vec{\hat{x}}.\vec{\hat{p}})
   +(\vec{\hat{p}}.\vec{\hat{x}})\vec{\hat{x}}\right].\vec{\hat{p}}\right\}
   \n\\
   &-&
   \frac{\lambda_{0}}{2}\left[\vec{\hat{x}}(\vec{\hat{x}}.\vec{\hat{p}})
   +(\vec{\hat{x}}.\vec{\hat{p}})\vec{\hat{x}}\right]^{2}.
 \eea

 \section{Transition golden rule}

Let us  now investigate briefly  the question of transition probabilities per unit
time between two arbitrary  states. Consider the
Hamiltonian $\hat{H}$, which is defined by equation (\ref{ht}) and  has  two parts $\hat{H}_{0}$ and $\hat{V}$. The
eigenvalues and eigenstates of $\hat{H}_{0}$ are also given in
 \cite{Higgs}.
%  \bea
%    |n\rangle
%    &=&
%    ,\n\\
%    E_{n}
%    &=&
%    (1+\frac{\lambda_{0}^{2}}{4})^{\frac{1}{2}}\
%(n+1)\hbar+\frac{\lambda_{0}}{2}\ \hbar(n+1)^{2}.
%  \eea
 We calculate the transition probability at time $t$  from an initial eigenstate $|i \rangle$ of $\hat{H}_{0}$
 to a final state $|j \rangle$ by treating  $\hat{V}$ as a perturbation.
 %%\be
 %%  P_{i\rightarrow j}(t)=\left|\langle j|\hat{U}(t,0)|i\rangle\right|^{2},
 %%\ee
 %%where $\hat{U}(t,0)$ is the evolution operator from time $0$ to $t$.
 %%Transition probability up to first order of perturbation
 %%is obtained as,
  \be
    P_{i\rightarrow j\neq i}(t)
    =\left|\frac{-i}{\hbar}\int_{0}^{t}e^{i \omega_{ji} t^{\prime}}
    V_{ji}(t^{\prime})d t^{\prime}
    \right|^{2},
  \ee
where, $\omega_{ji}=\frac{E_{j}-E_{i}}{\hbar}$ and,
 \be
   V_{ji}(t)=\langle j|\hat{V}(t)|i \rangle.
 \ee
Using  (\ref{vt}), the transition probability takes the following
form,
 \bea
   P_{i\rightarrow j\neq i}(t)
   =
   \frac{1}{\hbar^{2}}
   \left|\ V_{1_{ji}}
   \int_{0}^{t}e^{i \omega_{ji} t^{\prime}}
   V_{0}(t^{\prime};\alpha)d t^{\prime}
   +
   \widetilde{V}_{1_{ji}}
   \int_{0}^{t}e^{i \omega_{ji} t^{\prime}}
   \widetilde{V}_{0}(t^{\prime};\alpha)d t^{\prime}
   \right|^{2},
 \eea
where,
 \be
   V_{1_{ji}}=
   \langle j|\hat{V_{1}}(\vec{\hat{x}},\vec{\hat{p}};\lambda_{0})|i
   \rangle,
  \ee
and
 \be
   \widetilde{V}_{1_{ji}}=
   \langle j|\hat{\widetilde{V}_{1}}(\vec{\hat{x}},\vec{\hat{p}};\lambda_{0})|i \rangle.
 \ee
Using equation (\ref{v0}), we obtain,
 \bea
   \int_{0}^{t}e^{i \omega_{ji} t^{\prime}}
   V_{0}(t^{\prime};\alpha)d t^{\prime}
   &=&
   (\lambda_{0})^{\frac{1}{2}}\sum_{n}\alpha_{n}
   \omega_{n}
   \int_{0}^{t}e^{i \omega_{ji} t^{\prime}}\cos(\omega_{n}t)d
   t^{\prime}\n\\
   &=&
   -\frac{i}{2}(\lambda_{0})^{\frac{1}{2}}\sum_{n}\alpha_{n}
   \omega_{n}[\frac{e^{i(\omega_{ji}+\omega_{n})t}-1}{\omega_{ji}+\omega_{n}}
   +\frac{e^{i(\omega_{ji}-\omega_{n})t}-1}{\omega_{ji}-\omega_{n}}]\n\\
   &=&
   \frac{1}{2}(\lambda_{0})^{\frac{1}{2}}\sum_{n}\omega_{n}[A_{n}^{+}(t)+A_{n}^{-}(t)],
   \eea
where,
 \be
  A_{n}^{\pm}(t)=
  \alpha_{n}
  e^{[i(\omega_{ji}\pm\omega_{n})t/2]}\ \frac{\sin[(\omega_{ji}\pm\omega_{n})t/2]}{[(\omega_{ji}\pm\omega_{n})/2]}.
 \ee
We can also obtain by using equation (\ref{vp0}),
 \bea
   \int_{0}^{t}e^{i \omega_{ji} t^{\prime}}
   \widetilde{V}_{0}(t^{\prime};\alpha)d t^{\prime}
   &=&
   (\lambda_{0})^{\frac{1}{2}}\sum_{n}\alpha_{n}
   \int_{0}^{t}e^{i \omega_{ji} t^{\prime}}\sin(\omega_{n}t)d
   t^{\prime}\n\\
   &=&
   \frac{1}{2 i}(\lambda_{0})^{\frac{1}{2}}\sum_{n}\ [A_{n}^{+}(t)-A_{n}^{-}(t)].
 \eea
Now, to obtain the transition probability, we must calculate the
following term,
 \bea\label{sum}
   \left|\sum_{n} A_{n}^{+}(t)\left[\omega_{n} V_{1_{ji}}
   -i \widetilde{V}_{1_{ji}}\right] +
   \sum_{n} A_{n}^{-}(t)\left[\omega_{n} V_{1_{ji}}
   +i \widetilde{V}_{1_{ji}}\right]\ \right|^{2}.
 \eea
which contains the following three terms,
 \be\label{sum1}
   \left|\sum_{n} A_{n}^{+}(t)\left[\omega_{n} V_{1_{ji}}
   -i \widetilde{V}_{1_{ji}}\right] \right|^{2},
 \ee
 \be\label{sum2}
   \left|\sum_{n} A_{n}^{-}(t)\left[\omega_{n} V_{1_{ji}}
   +i \widetilde{V}_{1_{ji}}\right] \right|^{2},\\
 \ee
and
 \be\label{sum3}
   \sum_{n,m}
   A_{n}^{+}(t) A_{m}^{-}(t)^{\ast} \left[\omega_{n} V_{1_{ji}}
   -i \widetilde{V}_{1_{ji}}\right]
   \left[\omega_{m} V_{1_{ji}}
   -i \widetilde{V}_{1_{ji}}\right] +C. C.
 \ee
We can write equations (\ref{sum1}) and (\ref{sum2}) as sum of
 \bea\label{asl}
   \sum_{n} \left|A_{n}^{\pm}(t)\right|^{2}
   \left|\omega_{n} V_{1_{ji}}
   \mp i \widetilde{V}_{1_{ji}} \right|^{2}
 \eea
 and
  \bea\label{zee}
   \sum_{n,m} A_{n}^{\pm}(t) A_{m}^{\pm}(t)^{\ast}
   \left[\omega_{n} V_{1_{ji}}
   \mp i \widetilde{V}_{1_{ji}}\right]
   \left[\omega_{m} V_{1_{ji}}
   \pm i \widetilde{V}_{1_{ji}}\right]
   +C. C.
 \eea

It is clear from the equation (\ref{sum}) that if $t$ is large
enough, the probability of finding the system in the state
$|j\rangle$ will only be appreciable if the denominator of one of
the terms is close to zero. Moreover, both denominators can not
simultaneously be close to zero. According to similar arguments in
the standard quantum mechanics books \cite{Sakurai}, a good
approximation is therefore to neglect the interference terms
(\ref{sum3}) and (\ref{zee}) in calculating the transition
probability. Thus, if $\omega_{ji}\simeq\pm\omega_{n}$, only
$n$'th term in (\ref{sum}) will have an appreciable magnitude and
we can write (\ref{sum}) as,
 \bea\label{asl}
   |A_{n}^{\pm}(t)|^{2}
   \ |\omega_{n} V_{1_{ji}}
   \mp i \widetilde{V}_{1_{ji}} |^{2}
 \eea
which for large values of $t$ tends to take the following form,
\bea
   &&
   2\pi t \alpha_{n}^{2}\ |\omega_{n} V_{1_{ji}}
   \mp i \widetilde{V}_{1_{ji}} |^{2}
   \ \delta(\omega_{ji}\pm\omega_{n})\n\\
   &&
   =
   2\pi \hbar t
   \alpha_{n}^{2}\ |\omega_{n} V_{1_{ji}}
   \mp i \widetilde{V}_{1_{ji}} |^{2}
   \ \delta(E_{j}-E_{i}\pm\hbar\omega_{n}).
 \eea
The transition probability per unit time will, therefore, be
given by,
 \bea
    \Gamma_{i\rightarrow j\neq i}(t)
    &=&
    \frac{2 \pi}{\hbar}\frac{\lambda_{0}}{4}
    \sum_{n}\alpha_{n}^{2}\{|\omega_{n} V_{1_{ji}}
    -i \widetilde{V}_{1_{ji}} |^{2}
    \ \delta(E_{j}-E_{i}+ \hbar\omega_{n})\n\\
    &+&
    |\omega_{n} V_{1_{ji}}
    + i \widetilde{V}_{1_{ji}} |^{2}
    \ \delta(E_{j}-E_{i}- \hbar\omega_{n})\},
 \eea
which is the desired golden rule.

To summarize, we obtain the appreciable transiton probability only
if $E_{j}\simeq  E_{i}-\hbar\omega_{n}$ (stimulated emission) or
$E_{j}\simeq E_{i}+\hbar\omega_{n}$ (absorption). The isotropic
oscillator on the sphere may, therefore, be exited from lower
states to upper ones by using suitable frequencies of the
fluctuating background.
 If the oscillator's states are taken to represent a multi-level
atomic system, the multi-mode radius fluctuations can be
interpreted as a classical multi-mode radiation which interacts
with the atomic system \cite{Vogel}, \cite{Scully}. Our results
are expected to be equally  valid for other quantum systems on
more general time-dependent backgrounds.

  \section{Summary and Concluding Remarks}\label{summary}

In this paper,  a two-dimensional isotropic harmonic oscillator
was investigated  on a sphere with a time-dependent radius. It
was  shown that  variations in the sphere radius could be
represented by a minimally  coupled Hamiltonian. As a realization
of the model,  a two-dimensional isotropic harmonic oscillator
was considered on a fluctuating background. A simple golden rule
was obtained for the transition probabilities per unit time
between energy levels. This method can be generalized to other
time-dependent maximally symmetric surfaces, such as hyperbolic
or de-Sitter spaces. Generalization to an arbitrary surface by
using the approach of the reference \cite{da Costa} is also
interesting. The authors believe that new experimental methods
can be used to study quantum systems on time-dependent
backgrounds with important results. For example, in reference
\cite{Batz}, authors experimentally studied the impact of
intrinsic and extrinsic curvature of space on the evolution of
light. It should be possible to investigate evaluation of light
by experimental methods on a time-dependent background in the
near future.

    \section{Acknowledgment}
    A. M. would like to thank Shahrekord University for financial support.

\ \  \
\  \
%==================================================================================================
 
 \vspace{1cm}

%==================================================================================================

\end{document}